\documentclass[aip,jcp,preprint,amsmath,amssymb,superscriptaddress,showpacs]{revtex4-1}

\usepackage{graphicx}
\usepackage{dcolumn}
\usepackage{bm}
\usepackage{color}
\usepackage{soul}
\usepackage{textcomp}
\usepackage{rotating}
\usepackage{setspace}
\usepackage[mathlines]{lineno}
\usepackage{textcomp}
\usepackage{mathcomp}
\usepackage{sidecap}

\usepackage{microtype}
\usepackage[breaklinks=true,hyperindex=true,
pdftitle={Direct Experimental Determination of Spectral Densities of Molecular Complexes},
colorlinks=true,pagebackref=false,citecolor=blue,plainpages=false,pdfpagelabels,
linkcolor=blue,urlcolor=blue]{hyperref}

\begin{document}

\title{
Direct Experimental  Determination of Spectral Densities of Molecular Complexes
}

\author{Leonardo A. Pach\'on}
\affiliation{Grupo de F\'isica At\'omica y Molecular, Instituto de F\'{\i}sica,  Facultad de Ciencias Exactas y Naturales, 
Universidad de Antioquia UdeA; Calle 70 No. 52-21, Medell\'in, Colombia}
\affiliation{Chemical Physics Theory Group, Department of Chemistry and
Center for Quantum Information and Quantum Control,
\\ University of Toronto, Toronto, Canada M5S 3H6}
\author{Paul Brumer}
\affiliation{Chemical Physics Theory Group, Department of Chemistry and
Center for Quantum Information and Quantum Control,
\\ University of Toronto, Toronto, Canada M5S 3H6}

\begin{abstract}
Determining the spectral density of a molecular system immersed in a proteomic
scaffold and in contact to a solvent is a fundamental challenge
in the coarse-grained description of, e.g., electron and energy transfer dynamics.
Once the spectral density is characterized, all the time scales are captured and
no artificial separation between fast and slow processes need be invoked.
Based on the fluorescence Stokes shift function,
we utilize a simple and robust strategy to
extract the spectral density of a number of molecular complexes from available experimental data.
Specifically, we show that experimental data for dye molecules in several solvents, amino
acid proteins in water, and some photochemical systems (e.g., rhodopsin and green
fluorescence proteins),  are  well described by a three-parameter family of sub-Ohmic
spectral densities that are characterized by a fast initial Gaussian-like decay followed by a slow algebraic-like decay rate at long times.
\end{abstract}

\date{\today}

\maketitle

\section{Introduction}
The accurate description and quantification of  solvent effects on the electronic dynamics
of molecular systems are of great importance to reactions,\cite{HTB90}
various spectroscopies\cite{FC96,Muk99} and to  charge and electronic energy transfer\cite{MK11}.
In these cases  solvent and, as we discuss below,   intra- and
inter-molecular vibrations  play a dual role: the solvent screens the
electronic dynamics due to the spectral broadening but at the same time  provides the
energy to initiate chemical reactions and to stabilize  products once  reactions have
occurred\cite{FC96}.
%

Despite  its relevance, and due to the number of degrees of freedom involved in the
description of such systems, the role of the solvent and vibrations (induced, e.g., by proteomic scaffolds or nuclear motion) are effectively treated by
means of statistical approaches \cite{BKC90,WP91,FC96,Muk99,IC&10,
MK11,PB11,MW&11,PB12,CP&13}.  In these  approaches, the fundamental ingredient is the spectral density, which  models the 
effect  of the solvent and vibrations.
Once the spectral density is characterized, \emph{all} the time scales involved in the
dynamics are captured and there is no need for an artificial separation between fast and
slow processes.
Moreover, having the exact spectral density for a particular system
allows for the inclusion of quantum correlations (see below) in the calculation of solvated processes (e.g.,
electron transfer)\cite{BKC90, WP91,FC96}.

Being a key element in the effective description of the system-bath  coupling, it is clear that the equilibrium and non-equilibrium features of the system
dynamics are heavily determined by the spectral density $J(\omega)$.
In order to gain information
about $J(\omega)$ it  suffices, in accord with the fluctuation dissipation theorem (see below)\cite{CW51},  to measure either the non-equilibrium
relaxation or the equilibrium fluctuations.
This connection lead Fleming and Cho\cite{FC96} to suggest 
a spin-boson approach that allows extracting the spectral density from the Stokes shift response-function.
Despite  the  available experimental data for various systems, e.g.,
dye molecules\cite{JF&94,HSM97}, amino acid proteins\cite{CM&02}, or photochemical
complexes\cite{KF&01,AC&07}, and the clear advantage of extracting
$J(\omega)$ directly from experiment, this robust strategy  remains  unexplored.
Only  recently, based on quantum-classical \emph{simulations} of the non-equilibrium features,
has  this connection been used for  light-harvesting systems,\cite{VEA12} leading to a  calculated 
highly structured spectral density for  FMO, which is in good agreement with the spectral density inferred from the absorption
spectrum\cite{AR06}.

We show below that, based on experimental data for the Stokes shift response-function,
the initial fast relaxation followed by a slow relaxation, typical in  available data\cite{JF&94,
CM&02,KF&01,AC&07}, can be well characterized by the three-parameter family of sub-Ohmic
spectral densities,\cite{LC&87,Wei12,KA13}  whose features are discussed below.
Sub-Ohmic spectral densities are typical for noisy process in solid state devices
at low temperatures such as superconducting qubits\cite{SMS02} and quantum dots\cite{TV06}.
They also appear in the context of ultra-slow glass dynamics\cite{RNO03}, quantum impurity
systems\cite{SR&01}, nanomechanical oscillators\cite{SGC07} and fractal environments
such as porous and viscoelastic media\cite{LC&87} and reservoirs with chaotic dynamics\cite{Wei12}.
An important feature of the parametric family of
sub-Ohmic spectral densities is the fact that, for a broad class
of members, incoherent relaxation does not occur.
Hence, coherent processes can have long lifetimes, even in the regime of
strong coupling to the environment\cite{KA13}.

This paper is organized as follows: the formalism that relates the Stokes shift 
response function $S(t)$ to $J(\omega)$ is described in Sect. II. Note is made of the primary requirement,
adherence to linear response theory, and justification cited for the
applicability of linear response for the cases studied. Section III discusses
attributes of the sub-ohmic spectral density, extracted in Sect. IV from
experimental data. Additional aspects of this approach are the subject of Sect. V and
the Appendices. Section VI contains a brief summary.

\section{The Stokes Shift Response-Function}
The Stokes shift response function (or time-dependent solvation correlation function)
describes the solvent response to a sudden change in the charge distribution of a solute
molecule,\cite{HSM97,MG01,JF04,ML&10} and has been measured over different
time scales and for a variety of polar solvents (cf. Refs.~\citenum{HSM97,MG01,JF04,
ML&10} and references therein).
We briefly rederive the relationship between the fluorescence Stokes shift and the spectral density
in order to expose the underlying assumptions.
The essential argument follows that in Ref. \citenum{FC96}.

\begin{figure}[h]
\includegraphics[width=0.5\columnwidth]{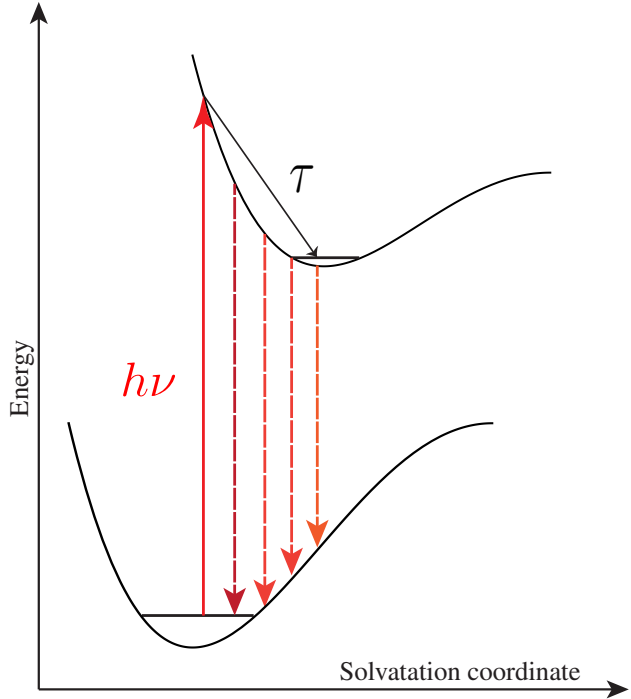}
\caption{Dynamic Stokes shift of a dipolar molecule in a polar environment}
\label{fig:ssoft}
\end{figure}

In typical Stokes shift experiments, a chromophore solute in a polar solvent is first excited
by a pump pulse, and  the time-dependent fluorescence spectrum of the solute is then
recorded\cite{HSM97} (see Fig.~\ref{fig:ssoft}).
In terms of experimental accessible quantities, the fluorescence Stokes shift function is defined
as\cite{FC96}
\begin{equation}
S(t) = \frac{\Delta E(t) - \Delta E(\infty)}{\Delta E(0) - \Delta E(\infty)},
\end{equation}
where $\Delta E(t)$ is the non-equilibrium energy difference between the excited state and
the ground state and is proportional to the time dependence of a characteristic fluorescence
frequency.
The goal is then to relate the non-equilibrium relaxation encoded in $S(t)$ with equilibrium
fluctuations of the energy difference and then extract the spectral density.
In doing so, let us a consider a two-electronic state system, with electronic transition frequency
$\omega_{\mathrm{eg}}$, coupled to a thermal bath $\hat{H}_{\mathrm{B}}$ via the interaction
term $\hat{V}_{\mathrm{SB}}$, and to a radiation field $\mathbf{E}(\mathbf{r},t)$ via the dipole
moment $\boldsymbol{\mu}$, so that the Hamiltonian is described by
\begin{equation}
\label{equ:Hamiltonian}
\hat{H} = \frac{1}{2}\hbar \omega_{\mathrm{eg}} \hat{\sigma}_z
+ \boldsymbol{\mu}\cdot\mathbf{E}(\mathbf{r},t)\hat{\sigma}_x
+\frac{1}{2} \hat{\sigma}_z \hat{V}_{\mathrm{SB}}(\mathbf{q},Q)
+ \hat{H}_{\mathrm{B}}(\mathbf{p},\mathbf{q},Q).
\end{equation}
Here $\hat{\sigma}$'s are the Pauli matrices and $\mathbf{q}$ and $Q$ denote the bath
and system coordinates, respectively.

Following Ref.~\citenum{FC96}, we use the Heisenberg picture and split the interaction
term $\hat{V}_{\mathrm{SB}}$ into an average and a fluctuating part,
$\delta V_{\mathrm{SB}}(t) = V_{\mathrm{SB}}(t) - \langle V_{\mathrm{SB}}\rangle$.
As is customary,  in order to have an effective description
of the coupling to the bath, one assumes that $\delta V_{\mathrm{SB}}(t)$ behaves as a random
variable and that it is characterized by its symmetrized,
$C_+(\tau') = \frac{1}{\hbar} \langle \{\delta V_{\mathrm{SB}}(\tau'), \delta V_{\mathrm{SB}}(0)\}\rangle$,
and anti-symmetrized correlation functions\cite{FC96,Muk99},
$C_-(\tau') = \frac{\mathrm{i}}{\hbar} \langle [\delta V_{\mathrm{SB}}(\tau'), \delta V_{\mathrm{SB}}(0)]\rangle$.
For convenience below we note that in accord with Kubo's formula, the linear response of the
system to the fluctuating perturbation $\delta V_{\mathrm{SB}}(t)$ is defined as the mean value
of the commutator $\delta V_{\mathrm{SB}}(t)$, i.e., in terms of the anti-symmetrized correlation
function $C_-(t)$.
The spectral density $J(\omega)$, the central element in the present discussion, is then defined
in terms of the Fourier transform of the response function, so that\cite{FC96}
\begin{equation}
\label{equ:defSpecDens}
J(\omega) = \frac{2}{\pi h} \frac{\Im \tilde{C}_-(\omega)}{\omega^2},
\end{equation}
where $\tilde{C}_-(\omega) = \int_0^{\infty} \mathrm{d}t \mathrm{e}^{\mathrm{i} \omega t} C_-(t).$

For  the Stokes shift function, one assumes that the energy difference operator can be
divided into two contributions
$
\Delta E(t) = \langle \Delta E(t) \rangle + \delta V_{\mathrm{SB}},
$
where $\langle \Delta E(t) \rangle $ is the average transition energy.
Linear response theory then allows the fluctuation due to the
perturbation to be generally written as an integral over the response function.
The normalized  fluorescence Stokes shift function $S(t)$ then becomes
$
S(t) = \int_t^{\infty} \mathrm{d}s\, C_-(s) / \int_0^{\infty} \mathrm{d}s\, C_-(s).
$
Given the relationship between the spectral density and the response function
[Eq.~\ref{equ:defSpecDens}], the fluorescence Stokes shift  function can be rewritten
in terms of the spectral density as
\begin{equation}
\label{equ:Soft}
S(t) = \frac{\hbar}{\lambda} \int_0^{\infty} \mathrm{d}\omega \omega
J(\omega) \cos \omega t,
 \end{equation}
where the normalization constant
$ \lambda = \hbar \int_0^{\infty} \mathrm{d}\omega \omega J(\omega)$
is identical to the solvent reorganization energy. It can be obtained experimentally from $S(t=0)$, or by any alternatively available route.

By inverting Eq.~\ref{equ:Soft}, the spectral density is obtained directly as
\begin{equation}
\label{equ:spectdens}
J(\omega) = \frac{1}{\pi}\frac{\lambda}{h \omega} \int_0^{\infty} {\rm d} t S(t) \cos(\omega t).
\end{equation}
This expression allows us, by means of a simple Fourier transform, to obtain the 
spectral density for a given physicochemical system from a measured $S(t)$.
The $J(\omega)$ so obtained is as accurate as is the observed $S(t)$, and provides a
direct means of understanding features of $S(t)$ in terms of the underlying $J(\omega)$.
Further, improvements in the measured $S(t)$ lead naturally to improved values for $J(\omega)$.
Note that alternative ways to estimate the spectral density rely on fitting procedures based on
the absorption and fluorescence spectra, which are obtained from the line shape function
[e.g., see Ref.~\citenum{Muk99}].
Thus, whereas alternative approaches provide only indirect information about the spectral
density, Eq.~\ref{equ:spectdens} offers a direct route between $J(\omega)$ and the measured data.

Before proceeding further, some comments are in order.
(i) For an harmonic bath, the right hand side of Eq.~\ref{equ:Soft} defines the damping kernel
$\gamma(t)$ discussed elsewhere\cite{PB12} and, within the
spin-boson case,   defines $S(t)$.
(ii) Note that  the Hamiltonian in Eq.~\ref{equ:Hamiltonian} implicitly assumes that the
coupling to the bath describes the interaction with the solvent and the  intra- and inter-nuclear
co-ordinates, i.e.,  that the Franck-Condon progressions include \emph{all} the
coupled modes.
(iii) The present analysis is based on the linear response approximation which 
 is expected to be accurate for chromophores with relative large size
and modest charge variations\cite{NH05}, such as those treated here.
It fails in significantly different types of systems, e.g.,
small solutes showing sizeable variations in atomic charges\cite{NH05}.
Thus,  we expect that an effective description
based on Eq.~\ref{equ:Soft} is sensible for the systems discussed below.  
(iv) Equation  (\ref{equ:spectdens}) makes clear that the higher the resolution of the observed $S(t)$, the more detail obtainable for $J(\omega)$.
  
  Below we develop this approach for a number of examples.

\section{Dye Molecules:  Motivating the Sub-Ohmic Spectral Density}
\begin{figure}[h]
	    \includegraphics[ width=0.5\columnwidth]{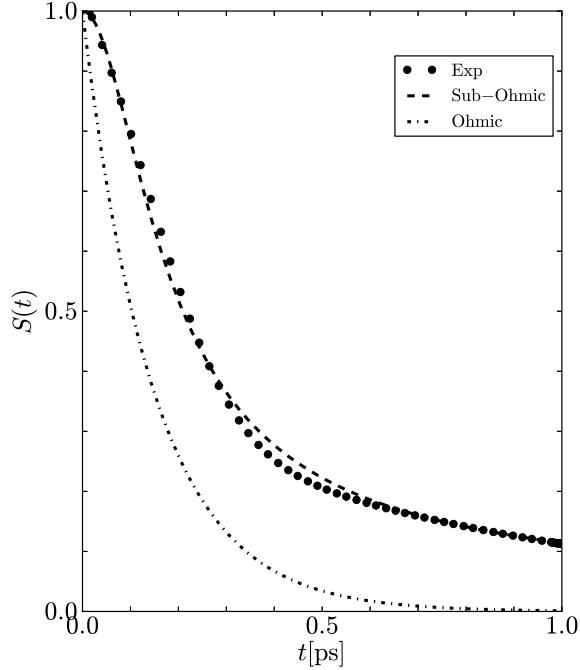}
  \caption{
               Dynamic fluorescence Stokes shift response function for the case of coumarin
               343 with water as a solvent.
	      }
\label{fig:soft}
\end{figure}

In Ref.~\citenum{JF&94}, the Stokes shift response function was measured for Coumarin 343, with water as a solvent.
The experimental results were well characterized by three terms, a fast Gaussian
decay and two slow exponential decays,
\begin{equation}
\label{equ:Sofexp}
S(t) = a_{\mathrm{g}} \mathrm{e}^{-\frac{1}{2}\omega_{\mathrm{d}}t^2}
+ a_1 \mathrm{e}^{-t/\tau_1} + a_2 \mathrm{e}^{-t/\tau_2},
\end{equation}
with $a_{\mathrm{g}} = 0.48$, $\omega_{\mathrm{d}} = 38.5$~ps$^{-1}$, $a_1 = 0.20$,
$\tau_1 = 0.126$~ps, $a_2 = 0.35$ and $\tau_2 = 0.880$~ps.
The time dependence  of $S(t)$ is depicted in Fig.~\ref{fig:soft}.
This figure shows a universal behavior present in many solvated experiments
in water [and, in general, in polar solvents such as acetonitrile or methyl chloride (cf.
Ref.~\citenum{FC96} and references therein)]\cite{JF&94,CM&02,NH05}: an  ultrafast Gaussian-type
 decay followed by an exponential-type slow component.
In this context, the ultrafast component is usually associated with librational (rotational) motions
of water molecules (or molecules of light molecular solvents) while the slow component, due to
the mass of the oxygen atom, is associated with translational motion\cite{JF&94}.
In the presence of a proteomic scaffold, the scaffold  vibrations  can also contribute to the
translational motion of water molecules\cite{NH05} and therefore contribute to the slow decaying
component.

Given  Eqs.~(\ref{equ:spectdens}) and (\ref{equ:Sofexp}), the Stokes shift response function, the spectral density follows
straightforwardly from Eq.~\ref{equ:spectdens} and, for this case, would be given by
\begin{align}
\label{equ:JomegaFitted}
\begin{split}
J(\omega) &\propto \sqrt{\frac{1}{2\pi\omega_{\mathrm{d}}}} \frac{a_{\mathrm{g}}}{\omega}
\mathrm{e}^{-\frac{1}{2 \omega_{\mathrm{d}}}\omega^2}
+ \frac{a_1 \tau_1}{\pi\omega(1 + \tau_1^2 \omega^2)}
+ \frac{a_2 \tau_2}{\pi \omega(1 + \tau_2^2 \omega^2)}.
\end{split}
\end{align}

Although the spectral density in Eq.~\ref{equ:JomegaFitted} or the Stokes function in Eq.~\ref{equ:Sofexp}
fit the experimental data, they contain six free parameters that are
essentially artificial because
one would expect, from the character of the data,
that only two gross time scales would suffice.
Hence, our proposal is to numerically evaluate Eq.~\ref{equ:spectdens} 
using the observed $S(t)$ and extract $J(\omega)$.  Prior to doing so, we recognize in
Fig.~\ref{fig:soft} a typical behavior well known in the context of open quantum systems\cite{LC&87,Wei12}:
the one induced by sub-Ohmic spectral densities. 
Hence,  we  first describe below the main properties of this parametric family of spectral
densities, and then make a connection with results for Coumarin 343,  amino acid proteins\cite{CM&02}, bovine
rhodopsin\cite{KF&01} and green fluorescence proteins such as mPlum, mRFP and mRaspberry\cite{AC&07}.
(The Appendix deals with some misunderstandings\cite{TS02}
that suggest that sub-Ohmic spectral densities are not acceptable.)

\subsection{Sub-Ohmic spectral densities}
\label{subsec:subOhmSpecDens}
An initial fast
relaxation followed by a slow relaxation is common in many solvated measurements.
This behavior is  familiar in noisy processes in superconducting qubits\cite{SMS02}
and quantum dots\cite{TV06}, ultra slow glass dynamics\cite{RNO03}, quantum impurity
systems\cite{SR&01}, nanomechanical oscillators\cite{SGC07} and fractal environments\cite{LC&87}
and is often described in terms of the parametric family of sub-Ohmic spectral densities  described by
\begin{equation}
\label{equ:subOhmicSD}
J(\omega) = 2 \delta_{s} \omega_{\mathrm{ph}}^{1-s} \omega^{s-2} \exp(-\omega/\omega_{\mathrm{c}}),
\end{equation}
with $0< s<1$.  The Ohmic spectral density follows from the case of $s=1$.
Here, $\delta_s$ is the dimensionless coupling-to-the-environment constant, $\omega_{\mathrm{c}}$ is
a cutoff frequency and $\omega_{\mathrm{ph}}$
is an auxiliary phononic scale frequency, not present in the Ohmic case, such that the
relevant coupling constant is $\delta_{s} \omega_{\mathrm{ph}}^{1-s}$.
The parameter values are determined
by the nature of the environment and its interaction with the system.
\footnote{
Note that in other contexts\cite{KA13}, the spectral density is defined without the factor
$1/\omega^2$ in Eq.~\ref{equ:defSpecDens} and therefore for the sub-Ohmic case
$J(\omega) = 2 \delta_{s} \omega_{\mathrm{ph}}^{1-s} \omega^{s}
\exp(-\omega/\omega_{\mathrm{c}})$.}

For the sub-Ohmic case, the Stokes shift function is
\begin{equation}
\label{equ:Softx}
S(t) = (1+t^2 \omega^2_{\mathrm{c}})^{-s/2} \cos[s \arctan(t \omega_{\mathrm{c}})],
\end{equation}
where we note that  \emph{only two time scales are present}.
In the short time regime, $\omega_{\mathrm{c}} t \ll 1$,
$S(t) \sim 1 -\frac{1}{2}s(1+s)(\omega_{\mathrm{c}} t)^2$,
which resembles the functional form of  a Gaussian decay at short times.
In the long time regime, $\omega_{\mathrm{c}} t\gg 1$,
$S(t) \sim \cos(\frac{\pi}{2} s) (\omega_{\mathrm{c}} t)^{-s}$, so that the long time decay is 
algebraic, $1/t^s$.  In this case, the reorganization energy, introduced in Eq.~\ref{equ:Soft} reads
\begin{equation}
\label{equ:reorgenergy}
\lambda = 2 \delta_s \Gamma(s)
(\omega_{\mathrm{c}}/\omega_{\mathrm{ph}})^{s-1} \hbar \omega_{\mathrm{c}},
\end{equation}
where $\Gamma(s)$ denotes the gamma function of $s$\cite{GR07}.
A full characterization of the dynamics and spectral quantities, requires consideration of other quantities, such as the  Huang-Rhys factor,  discussed
 in the Appendix.

For the sake of completeness, we present explicit results based on the sub-Ohmic spectral
density Eq.~\ref{equ:subOhmicSD} for the relevant quantities in echo spectroscopies.  Assuming that $\delta V_{\mathrm{SB}}(t)$ obeys a Gaussian statistics,      it is possible to express the third-order non-linear signals in terms of the line shape function $g(t)$,
\begin{equation}
\label{equ:goft}
g(t) = \frac{1}{2\hbar} \int_0^t  \mathrm{d} \tau \int_0^{\tau} \mathrm{d} \mathrm{\tau}'
\left[ C_+ (\tau') + \mathrm{i} C_-(\tau')\right],
\end{equation}
where
$C_+(\tau')$ and $C_-(\tau')$ are the symmetrized and anti-symmetrized correlation function,
 defined above\cite{FC96,Muk99,HI05}.
The real part of $g(t)$ describes the spectral broadening, whereas the imaginary
part is related to the fluorescence Stokes shift [cf. Eq.~\ref{equ:Soft}].

In terms of the spectral density, the line shape function $g(t)$ in Eq.~\ref{equ:goft} can be expressed
as\cite{FC96,Muk99}
\begin{equation}
\label{equ:goftALTdef}
\begin{split}
g(t) &= -\mathrm{i} \lambda t/\hbar
+ \mathrm{i} \int_0^{\infty}\mathrm{d}\omega J(\omega)\sin \omega t
\\
&+ 2\int_0^{\infty}\mathrm{d}\omega J(\omega) \frac{1}{\mathrm{e}^{\hbar \beta \omega}-1}
(1-\cos \omega t)
+ \int_0^{\infty}\mathrm{d}\omega J(\omega) (1-\cos \omega t),
\end{split}
\end{equation}
where we can identify the time-derivative of the second term of the first line with the
Stokes shift function and  the second line contains  the effects of the thermal environment and
 zero-point fluctuations. For the particular case of the sub-Ohmic spectral densities, we have
\begin{equation}
\begin{split}
g(t) &= -\mathrm{i} \lambda t/\hbar + 2\delta_s \Gamma(s-1)
\left( \omega_{\mathrm{c}}/\omega_{\mathrm{ph}} \right)^{s-1}
 \left\{
1-(1+\mathrm{i} \omega_{\mathrm{c}} t)^{1-s}
+ 2\kappa^{s-1} \zeta(s-1,1+\kappa)
\right.
\\
& -\left. \kappa^{s-1}
\left[\zeta\left(s-1,1+\kappa + \mathrm{i}t/\hbar \beta\right) \right. \right.
+\left.\left.
 \zeta\left(s-1,1+\kappa - \mathrm{i}t/\hbar \beta\right)  \right]
\right\},
\end{split}
\end{equation}
with $\kappa = 1/\hbar \beta \omega_{\mathrm{c}}$ and $\zeta(z,q)$ is the generalized
Riemann's zeta function  
\footnote{For all values of $z$ expect $z=1,2,3,\ldots$,
see Eq.~9.512 in Ref.~\citenum{GR07},
$\zeta(z,q) = -\frac{\Gamma(1-z)}{2\pi \mathrm{i}}
\int\limits^{0^+}_{\infty} \mathrm{d}\theta \frac{(-\theta)^{z-1}}{1- \mathrm{e}^{-\theta}}\mathrm{e}^{-q\theta}$.}.
%
Once the line shape function is known, the fluorescence and absorption spectra can be
obtained.\cite{Muk99}

\subsubsection{Cumarin 343}

In order to explore to what extent the sub-Ohmic description is quantitative for the Cumarin
343 example in Fig.~\ref{fig:soft}, we fit
the experimental data from Ref.~\citenum{JF&94} to a sub-Ohmic spectral density with
$\omega_{\mathrm{c}} = 6.25846$~ps$^{-1}$ and $s=0.785158$.
As shown in Fig.~\ref{fig:soft},  the sub-Ohmic spectral density, with only two parameters,
correctly describes the fast initial decay and the subsequent slow relaxation observed in the
experiment and fitted to $S(t)$ in Eq.~\ref{equ:Sofexp}.
Also shown for comparison, is the single exponential decaying Stokes shift function derived for an
Ohmic spectral density, $S(t)=\exp(-\omega_{\mathrm{c}} t)$, which clearly fails to
describe the experimental behavior.

In the following, we consider the nature of $J(\omega)$ for  more complex systems, such as amino acid proteins and photochemical systems.

\section{Analyzed Spectral Densities}
\subsection{Amino Acid Proteins}
\begin{figure}[h]
\includegraphics[width = 9cm]{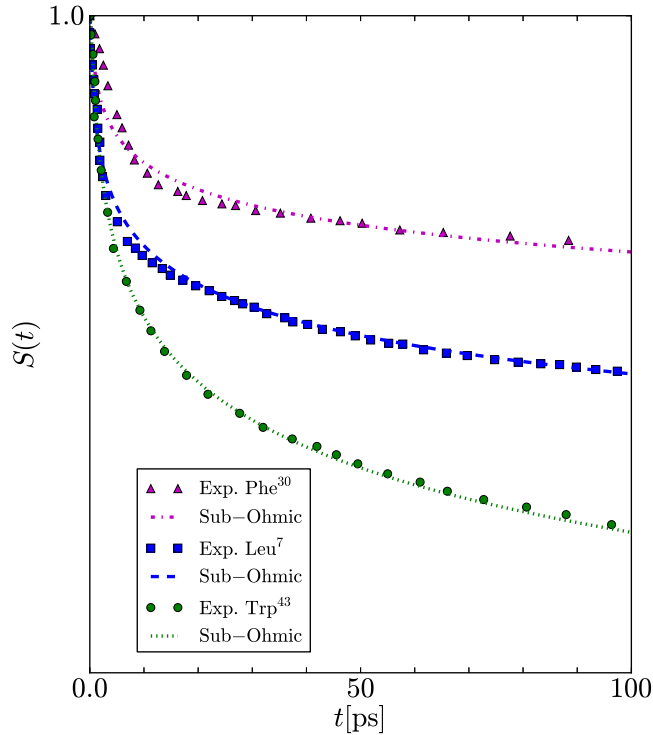}
\caption{Normalized dynamic Stokes shift, extracted from Ref.~\citenum{CM&02}, of GB1 Aladan
mutants as characterized by the time dependence of peak emission energies of the
Phe$^{30}$ (purple triangles with $\omega_{\mathrm{c}} = 1.59407$~ps$^{-1}$ and $s=0.003447$),
Leu$^7$ (blue squares with $\omega_{\mathrm{c}} = 5.8407$~ps$^{-1}$ and $s=0.00433494$) and
Trp$^{43}$ (green circles with $\omega_{\mathrm{c}} = 2.32145$~ps$^{-1}$ and $s=0.00735004$)
mutants.
}
\label{fig:softscience}
\end{figure}

For this particular case, as well as the case of pigment aggregates, an ideal probe for
studying protein dynamics and electrostatics should be sensitive to its environment and
should be able to be  incorporated, site-specifically, throughout any protein of interest\cite{CM&02}.
In Ref.~\citenum{CM&02}, Adalan, an environment-sensitive fluorescent amino acid,  was synthesized and site-specifically incorporated into proteins by both nonsense
suppression and solid-phase synthesis.
In particular, Adalan was used to probe the electrostatic character of the B1 domain of
streptococcal protein G (GB1) at multiple sites using time-resolved fluorescence.
Fig.~\ref{fig:softscience} shows the experimental results for the dynamic Stokes shifts for the Phe$^{30}$
(buried in the protein), Leu$^7$ (buried in the protein) and Trp$^{43}$ (partially exposed
in the protein)  mutants, extracted from Ref.~\citenum{CM&02} and our  fit using a sub-Ohmic
spectral density.
It is clear that the short as well as the long time dynamics are extremely well characterize by Stokes
shift function in Eq.~\ref{equ:Softx}, which is induced by the sub-Ohmic spectral density.

Understanding the origin  of the particular parameter  values from the  fitting procedure, which vary considerably for the three cases (see figure caption, Fig.~\ref{fig:soft})  requires
a deep analysis of the interaction between the pigments, the proteomic scaffold and the
solvent.
However, an immediate consequence of  our effective description is that it simplifies the calculation
of cross-grained quantities such as line shape functions described above  (see \ref{subsec:subOhmSpecDens}) and
excitation energy transfer rates [cf. Chap.~9 in Ref.~\citenum{MK11}] and provides a starting point for analysis of the features of the spectral density.

\subsection{Photochemical Systems}
\label{sec:SouDissDec}

Recently, there has been a great interest in the dynamics of photochemical systems
\cite{KF&01,PN&06,FC&09,AB13,PYB13,WP91,AC&07,IC&10,MK11,PB11,PB12,CP&13}.
In particular, two cases  have been  extensively investigated: (i) the \emph{cis/trans} isomerization
of rhodopsin,\cite{KF&01,PN&06,FC&09,AB13,PYB13}
and (ii) the energy transfer processes in natural light-harvesting systems.\cite{WP91,
AC&07,IC&10,MK11,PB11,PB12,CP&13}
Below, we explore the extent to which the sub-Ohmic spectral density describes
available experimental data for the dynamic Stokes shift of these systems.

\subsubsection{Rhodopsin}
\label{sec:PhotoRhod}
\begin{figure}[h]
\includegraphics[width = 10cm]{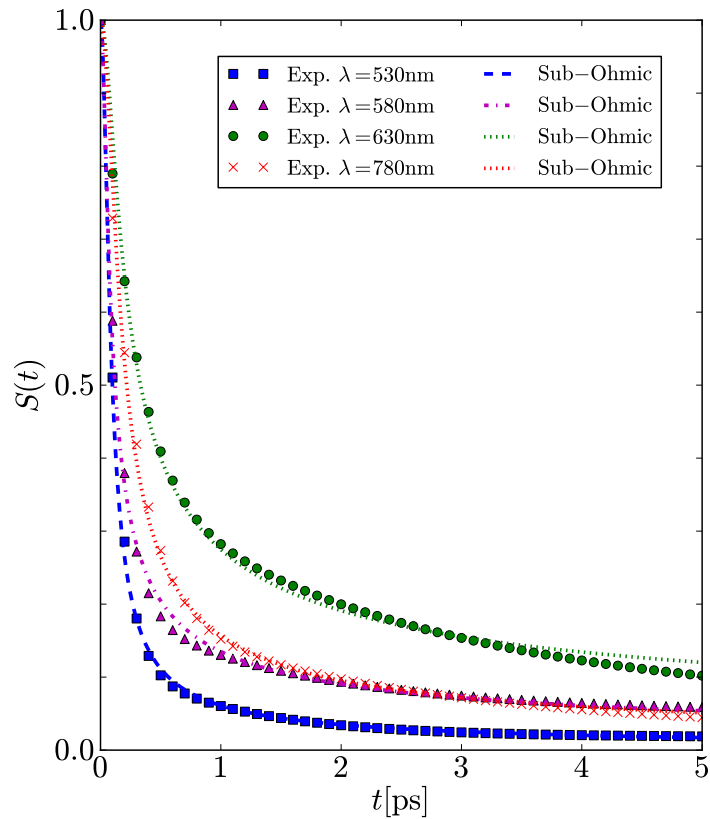}
\caption{Normalized dynamic Stokes shifts, extracted from Ref.~\citenum{KF&01}, of
bovine rhodopsin at various wavelengths:
$\lambda=530$~nm (blue squares, with $\omega_{\mathrm{c}} = 14.661$~ps$^{-1}$
and $s=0.736$),  $\lambda=580$~nm (purple triangles, with $\omega_{\mathrm{c}} = 17.878$~ps$^{-1}$ and $s=0.554$),
$\lambda=630$~nm (green circles, with $\omega_{\mathrm{c}} = 7.926$~ps$^{-1}$ and $s=0.489$),
$\lambda=680$~nm ($\omega_{\mathrm{c}} = 5.677$~ps$^{-1}$ and $s=0.529$ not shown),
$\lambda=730$~nm ($\omega_{\mathrm{c}} = 5.382$~ps$^{-1}$ and $s=0.594$ not shown) and
$\lambda=780$~nm (red \textsf{x}'s with $\omega_{\mathrm{c}} = 7.985$~ps$^{-1}$ and $s=0.643$).
}
\label{fig:softBR}
\end{figure}

Rhodopsin is an excellent molecular switch which converts light signals to the electrical
response of the photoreceptor cells,\cite{KF&01} and which has been extensively studied.\cite{KF&01,PN&06,FC&09}
In Ref.~\citenum{KF&01}, the Stokes shift function was measured for the case of bovine
rhodopsin at various wavelengths.
As shown in Fig.~\ref{fig:softBR}, the description based on the sub-Ohmic
spectral density is highly accurate in all cases, which correspond to different excitation wavelengths $\lambda$.
We note that the spectral density needs to be recalculated for each $\lambda$ since
each wavelength excites different constituents of the molecular
complex.
As a consequence, each wavelength induces  different behavior of the vibrational modes and solvent
response.

Note that this $J(\omega)$ representation  allows for a dramatic simplification of  calculations needed to, for example,  explore the time
evolution of rhodopsin since the effect of all Raman modes can now be \emph{effectively}
condensed in a sub-Ohmic spectral density.

\subsubsection{Green Fluorescent Protein}
\label{sec:GreenFluoProt}
Green Fluorescent Proteins (GFPs) are thought to be ideal candidates for measurements
of the dynamic Stokes shift because the chromophore is both intrinsic to the protein and structurally
well characterized\cite{AC&07}.
This feature was exploited in Ref.~\citenum{AC&07} to measure the dynamic Stokes shift
function in variants of GFP such as mPlum, mRFP and mRaspberry at different pH levels.
The experimental results available from Ref.~\citenum{AC&07} are those from a fitting
procedure to a log-normal line-shape with, due to the weak solvation, an added baseline parameter.
Hence, in order to properly describe the $S(t)$ based on the sub-Ohmic spectral density,
we also introduce a baseline contribution $b_0$, so that $S(t)$ for this case is
$S_{b_0}(t) = [S(t) + b_0]/(1+b_0)$, with $S(t)$ given by Eq.~\ref{equ:Softx}.

\begin{figure}[h]
\includegraphics[width = 9cm]{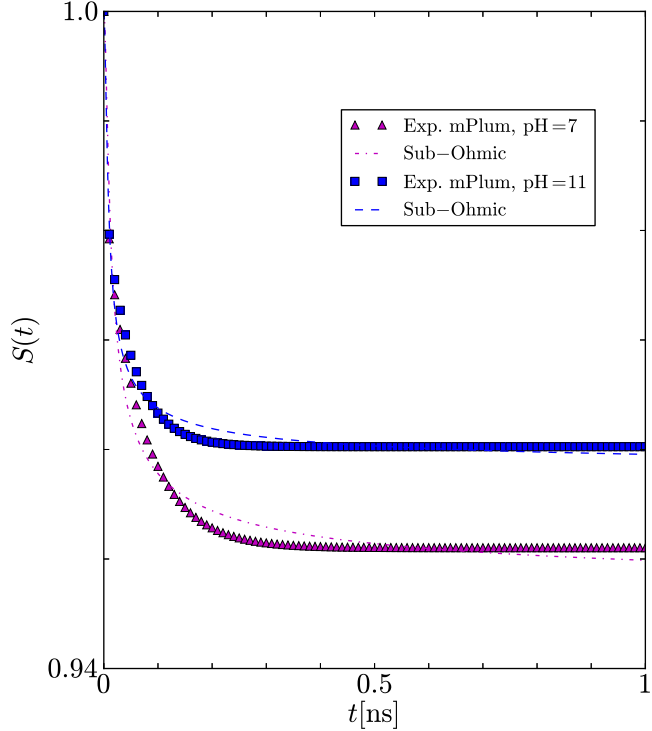}
\caption{Normalized dynamic Stokes shifts, extracted from Ref.~\citenum{AC&07} of mPlum
at different pH levels: pH= 7 (purple triangles, with $\omega_{\mathrm{c}} = 143.90$~ps$^{-1}$, $s=0.467$ and
$b_0=17.493\times10^3$~cm$^{-1}$), and pH = 11 (blue squares with $\omega_{\mathrm{c}} = 202.04$~ps$^{-1}$
and $s=0.5296$ and $b_0 = 22.722\times10^3$~cm$^{-1}$).}
\label{fig:softPNAS}
\end{figure}

In Fig.~\ref{fig:softPNAS}, we depict the experimental results and their characterization based on the sub-Ohmic
spectral density for mPlum at pH 7 and pH 11.
As in previous cases, the accuracy of the description is very good.
For the case of mRFP ($\omega_{\mathrm{c}} = 0.1995$~ps$^{-1}$
and $s=0.0011$) and mRaspberry ($\omega_{\mathrm{c}} = 0.1967$~ps$^{-1}$
and $s=0.0011$) in pH 7 buffer (not shown), the description is also  very accurate, with no
need for an additional baseline contribution.

\section{Additional Remarks}
We have discussed an experimentally accessible way to directly determine the spectral
densities of molecular complexes, which is also applicable  to systems in solid state
physics such as quantum dots.
Below, we discuss some technical points related to this formulation, and to semiclassical approaches aimed at calculating $S(t)$.

\emph{Classical vs. Quantum Correlations}\textemdash
Due to the sheer complexity of \textit{ab initio} calculations of the dynamics of large
physicochemical systems [cf. Ref.~\citenum{WP91,NH05,VEA12}],  computational studies of $S(t)$ usually
invoke  classical treatments of the nuclear dynamics.
Thus, in the classical limit, since the spectral distribution of
fluctuations is far less than $2k_{\mathrm B}T$, the response function is directly related to the
classical correlation function
$
C_{-\mathrm{cl}}(t) = -\beta \frac{\mathrm{d}}{\mathrm{d} t}C_{\mathrm{cl}}(t) =
-\beta \frac{\mathrm{d}}{\mathrm{d} t} \langle \delta V_{\mathrm{SB}}(t)
\delta V_{\mathrm{SB}}(0)\rangle_{\mathrm{cl}},
$
where the average is taken over classical phase space and $\delta V_{\mathrm{SB}}$ is a
classical variable.
Hence,
\begin{equation}
S_{\mathrm{cl}}(t) = \frac{\langle \delta V_{\mathrm{SB}}(t)\delta V_{\mathrm{SB}}(0)\rangle_{\mathrm{cl}}}
{\langle \delta V_{\mathrm{SB}}^2\rangle_{\mathrm{cl}}},
\end{equation}
which is a formal expression of the classical Onsager regression hypothesis.

It is important to note that, formally, the regression hypothesis fails in the quantum regime\cite{FO96}
and in the case of Markovian dynamics, violates the Kubo-Martin-Schwinger\cite{Kub57,MS59}
principle of detailed balance\cite{Tal86}.
That is,\cite{FC96} in the classical limit the temperature is the only parameter
determining the fluctuation-dissipation relation (cf. Ref\citenum{HI05} and references therein),
whereas in the quantum case, the fluctuation-dissipation relation requires  complete knowledge
of the spectral distribution of the fluctuation\cite{PT&14}., i.e., $J(\omega)$.
The present approach is based on experimental data, since it  has the advantage of making no reference
to the classical or quantum nature of the correlations, it just makes use of the  general
quantum formulation.   The ``quantumness" of the correlations then emerges directly from
the experimental data.
Thus, having the possibility of extracting the  spectral density directly from experimental
data is not  only of  practical importance, but also is significant  from a fundamental
viewpoint.

\section{Summary}

The results of this work show clearly that sub-Ohmic spectral densities provide an
excellent description of the spectral densities $J(\omega)$ associated with an impressive 
number of large solvated systems, and that such spectral densities can be
directly extracted, within the spin-boson model, from the experimental Stokes shift response function $S(t)$. 
Higher resolution $S(t)$  may yield more detailed $J(\omega$), but the smooth
underlying sub-Ohmic spectral density structure is expected to persist. These
results, plus 
the extensive literature on sub-Ohmic spectral densities that describe their physical
consequences and characteristics, 
encourage greater theoretical and experimental studies on both the Stokes shift response
function and on the interpretation of the observed sub-Ohmic parameters in physicochemical
systems. 

One further note is in order. 
Our recent work on the theory of
one-photon phase control of molecular systems\cite{SAB10,AB13,PYB13}, shows that the
sub-Ohmic character could assist the one-photon phase control, and effect that was motivated
by experiments
at low\cite{PN&06} and high\cite{FC&09} field intensities. As a consequence, the 
broad range of systems displaying sub-Ohmic behavior is encouraging for one-photon
phase control of such systems.

\acknowledgements
Comments by G. D. Scholes and references suggested by B. E. Cohen are gratefully acknowledged.
This work was supported by the US Air Force Office of Scientific Research under contract
number FA9550-13-1-0005, by the \emph{Comit\'e para el Desarrollo de la Investigaci\'on}
--CODI-- of Universidad de Antioquia, Colombia under contract number E01651
and under the \emph{Estrategia de Sostenibilidad 2013-2014}, by the \emph{Departamento Administrativo
de Ciencia, Tecnolog\'ia e Innovaci\'on} --COLCIENCIAS-- of Colombia under the grant number
111556934912.

\section{Appendix}

It is illustrative to examine  some previous considerations\cite{TS02} about
the zero phonon line  in absorption spectra of chromophores in solid structure.
There, it was argued that physically relevant spectral densities should converge to zero as
the frequency approaches zero.
For the widely used family of spectral densities of the form\cite{LC&87,Wei12}
$J(\omega) \sim \omega^{s-2}$, this would require  $s>2$, ruling out the sub-Ohmic
($0<s<1$), Ohmic ($s=1$) and super Ohmic ($1<s\le2$) spectral densities.
However, such as an argument arises from an incomplete understanding of the \emph{quantum}
fluctuation-dissipation theorem\cite{CW51} (see e.g. Ref.~\citenum{HI05} or Chap.~6 in
Ref.~\citenum{Wei12})  and the homodyne nature of the spectrum measurement-process
via absorption\cite{GZ04}.
Hence, it is  pertinent to comment on these issues.

\textit{The Quantum Fluctuation-Dissipation Theorem}.
The classical fluctuation-dissipation theorem formulated in 1928 by Nyquist\cite{Nyq28},
and experimentally verified by Johnson\cite{Joh28}, states that a resistor $R$, in response
to the inherent fluctuations in maintaining a Boltzmann distribution of the canonical variables
in an electric circuit, develops a current voltage $V(t)$ across its ends.
The Fourier-transformed two-point-correlation-function of the induced current,
$\mathcal{S}(\omega) = \frac{1}{2}\int_{-\infty}^{\infty} \mathrm{d}\tau
\,\mathrm{e}^{-\mathrm{i} \omega \tau}\langle V(\tau) V(0) \rangle$, is given by\cite{GZ04}
%
\begin{equation}
\mathcal{S}(\omega) = R k_{\mathrm{B}} T.
\end{equation}
Motivated by the work of Planck on the quantized spectrum of the blackbody radiation,
in the last paragraph of his 1928 paper, Nyquist considered the case when
$\hbar \omega > k_{\mathrm{B}} T$, which he suggested should be equivalent to taking
$\mathcal{S}(\omega) = 2 R \hbar \omega
\left[1/\exp(\hbar \omega/  k_{\mathrm{B}} T) -1 \right]^{-1}$.

A formal treatment of the quantum  results of Nyquist was provided by Callen and
Welton,\cite{CW51} who showed that the correct correlation function $\mathcal{S}(\omega)$
reads
\begin{equation}
\label{equ:quantspect}
\mathcal{S}(\omega) = 2\left[ \frac{1}{2}\hbar \omega
+ \frac{\hbar \omega}{\exp(\hbar \omega/  k_{\mathrm{B}} T) -1} \right] \Re Y(\omega),
\end{equation}
where $Y(\omega)$ is equivalent to the susceptibility, which for the case of a Markovian
Ohmic resistor corresponds to $R$, i.e.,  frequency independent dissipation.
Thus, by contrast to Nyquist's original suggestion, at low temperature
($k_{\mathrm{B}} T / \hbar \omega \rightarrow 0$), the $\mathcal{S}(\omega)$ grows linearly
with $\omega$ due to the zero point fluctuations , i.e. due to the quantum noise.
Most importantly, the quantum noise has an associated non-zero susceptibility and
therefore one expects  a finite width to the zero phonon line (ZPL).
However, as the authors in Ref.~\citenum{TS02} correctly noted, for a multitude of
molecular systems there is no experimental evidence  of such a finite width for the
ZPL at $T=0$~K.
This observation lead them to argue,  incorrectly, that the ZPL profile has zero
width at 0~K.
\vspace{0.5cm}

\textit{The Homodyne Nature of the Spectrum Measurement-Process by Absorption}.
The fact that the experimental data does not provide any evidence of a finite width of the
ZPL at $T=0$~K is analogous to the fact that the detected blackbody-radiation-spectrum
is well described by the Planck distribution with no contribution of the linear term in
Eq.~\ref{equ:quantspect}.
The reason for this is that we cannot detect the zero point energy contribution to the
spectrum when we use experimental homodyne schemes that are based on the absorption of
photons from a radiation field\cite{GZ04}.
Specifically,  in homodyne absorption measurements, what is measured
is the normal product of the creation and annihilation operators of the field (see, e.g.,
Chap.~8 in Ref.~\citenum{GZ04}), and hence there is no zero point contribution.
However, this does not mean that the effect of the zero point energy is not accessible
experimentally; it was indeed measured\cite{KVC82} by means of heterodyne detection
at 1.6 and 4.2~K.
However, at high temperature $\hbar \omega / k_{\mathrm{B}} T > 1$, the main contribution
comes from the Planck distribution of the thermal bath, and the spectral lines are expected
to have a finite width.

\emph{On the Huang-Rhys factor for Sub-Ohmic Spectral Densities}.
The Huang-Rhys factor can be seen as a measure of the effective mass
of the environment.\cite{GSI87,Wei12}  For the case of the sub-Ohmic spectral density
introduced in Eq.~\ref{equ:subOhmicSD} it is defined as
\begin{equation}
\mathcal{S}_{\mathrm{H-R}} = 2 \delta_{s} \omega_{\mathrm{ph}}^{1-s}
\int_0^{\infty} \mathrm{d} \omega \frac{\omega^{s}}{\omega^2} \exp(-\omega/\omega_{\mathrm{c}}).
\end{equation}
In the context of open quantum systems\cite{Wei12}, the divergence of this integral is known
to lead to orthogonality catastrophe, a concept we will not discuss here.
In the super-Ohmic case $s>1$, the integral in infrared-convergent and, in the case of a
two-level system,  is an indication of possible elastic tunneling without dynamical involvement
of the bath.
In the sub-Ohmic case $0<s<1$, the integral is infrared-divergent, which means that the
low frequency modes must be treated non-adiabatically, a procedure that can be found
in  Ref.~\citenum{Wei12}, Sec.~20.2.



%

\end{document}